# A Ғlexible FPGA-Based ISA Configurable SoC platform

Shih-Yi Yuan [1,*], Bo-Yu Zhu [2,*]

1. Shih-Yi Yuan, Department of Communication Engineering, Feng Chia University, 40724 Taichung, Taiwan
2. Bo-Yu Zhu, Department of Communication Engineering, Feng Chia University, 40724 Taichung, Taiwan
* Correspondence: yuanmark@gmail.com(S.Y.); chu19961221@gmail.com(B.Z.)

**Abstract:** We proposes a platform which can generate hardware/software description based on flexible instruction set architectures (ISAs). The platform takes advantage of the flexibility of field programmable gate array (FPGA) to design many micro control units (MCUs) based on different ISAs. The platform can generate many ISAs, MCUs, and Assemblers according to a pre-defined ISA and user applications. Although the MCU performance is not optimized, the FPGA shows a great potential on resource reduction and enough performance at very low system clock rate. The flexible ISA has shown great importance for the design targeted to specific purpose. We also show a case study of the proposed flexible ISA-based FPGA-MCU. It can control many specifically designed hardware IPs and a customized multi-task OS with tasks. Not only the case works correctly, but also the proposed FPGA-MCU of the case is flexible with reduced FPGA resources, low cost, and within time constraints.

## 1. Introduction

Nowadays, CPU generally exist in every kind of electronic devices. It can execute many programs sequentially or parallelly. Programs are used to extend functionalities of a CPU. However, it is still limited by CPU instruction set. If a CPU wants to control a special hardware function, the direct way is to add the dedicated hardware function and add an instruction to instruction set architecture (ISA) which can control the hardware. However, the general-purpose micro control unit (MCU) or CPU cannot be customized easily. They are difficult to add or delete hardware functions or instructions. The reason is that despite the hardware functions and ISAs should be modified, the tool chains must be modified accordingly. It includes both hardware and software modifications and are difficult. Thus, the performance or resource cannot be further improved when the MCU/CPU ISA is fixed.

The main features of field programmable gate array (FPGA) are flexibility, integration, and cost saving [1]. Recent years, the application of FPGA devices has increased in a wide variety of fields[2], such as: digital signal processing [3–7], data processing [8,9], bioinformatics [10,11] and power electronics [12–14]. Compared with CPU, FPGA can be designed more flexibly and can improve the final application's performance with less resources [15]. For example, FPGA can be used to develop its own dedicated CPU, MCU, or SoC system [16-19]. The system can remain the convenience of a CPU system while gaining the flexibility and security benefit from FPGA. It can be designed to match with target application requirements such as customized bus protocols, different numbers and functions of specialized peripheral input/output (I/O), or protected IPs.

Generally, when an ISA of a specific CPU is fixed, toolchain engineers can design various programming tools (assembler, compiler, linker, loader, …) to enhance the convenience of application programmers. However, toolchain design and optimization are a highly skilled field, which is difficult for application designers to design by themselves in a short time. Because of the flexibility and variability of the hardware architecture of FPGAs, the design of the toolchain must rely on toolchain designers to design them first. In the meantime, application designers have to wait until such tools are available. This limits the development of FPGA.

In response to this problem, this paper proposes a platform which can automatically generate FPGA design descriptions based on ISA descriptions and semi-automatically generate toolchain tools. The platform has the advantages described below.

In this paper, a case study of system on chip (SoC) is designed based on the proposed platform. The SoC includes a specially designed ISA, an auto-generated CPU based on the ISA, a bus controller, and various peripheral devices (GPIOs, an UART, and Timers). Although the cost of the proposed CPU cannot be compared with MCUs on market [20], the flexibility of the FPGA design platform can be very beneficial for dedicated users based on their requirements, considering the flexibility and performance with dedicated hardware functions. And we use the case study to justify the claims.

Due to the limitation of current FPGA logic resources, we have simplified the complexity of the CPU to demonstrate the scenario we proposed. The FPGA-based CPU is simplified as much as possible to demo the mechanism that can automatically generate a hardware and an assembler with a minimum number of instructions for the application that the fulfills the user requirements. During the design phase, redundant functional blocks are automatically removed thereby minimizing FPGA's resource usage. Considering the impact of ISA changes upon the toolchain, we will design software tools that can reduce the ISA (R-ISA). At the same time, the tool chain uses the R-ISA to generate a new assembler. At present, the toolchain only includes an assembler and a linker. It will be expanded to other tools such as compiler, loader, etc. in the future.

**2. The Proposed Platform**

A.  ISA of the target CPU design

The CPU-ISA used in this paper is based on a RISC-type ISA [21] adapted from CPU032I ISA [22]. There are totally 36 instructions, and no floating-point instructions. The CPU architecture is shown in Figure 1. Other ISAs can be inserted or modified through the platform.

Current CPU (Figure 1) doesn't have pipeline. Its instruction cycle (Fetch, Decode, Execute, Write-back) needs four system clocks. Current register unit contains 17x32-bit registers. R0 is a zero register, R1~R11 are general registers, R12 is a status register (SW), R13 is stack pointer (SP), R14 is the link register (LR), R15 is the program counter (PC), and R16 is the instruction register (IR). The control unit (CU) is responsible for controlling instruction cycles of the CPU (fetching, decoding, and execution). Through a bus controller, the CPU can access specified IPs, memory, or I/O units.

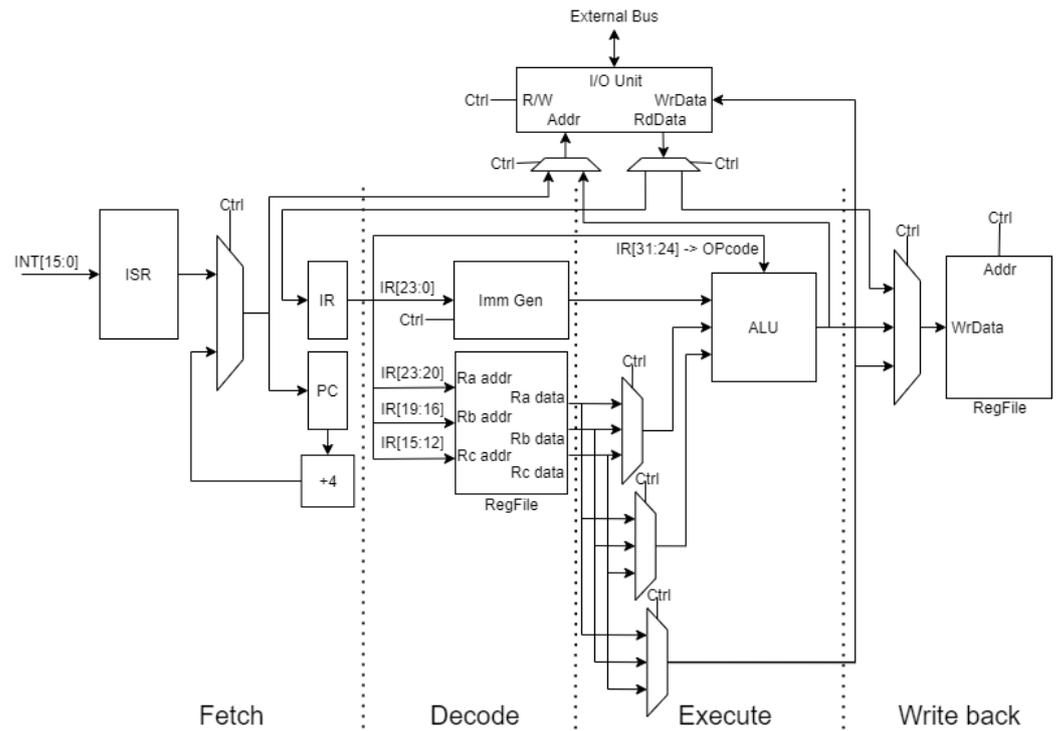

**Figure 1.** The proposed 32 bits CPU basic architecture (based on [21]).

B. Different IPs for SoC platform

CPU must be paired with other external devices to perform its function. So we designed a SoC in FPGA. Taking the advantage of FPGA flexibility, we can design any functional circuit flexibly. We also adapted several external devices (GPIO, UART, Timer) connecting to the Bus. And we provide a set of demonstrations to illustrate the concept. The demonstration architecture is shown in Figure 2.

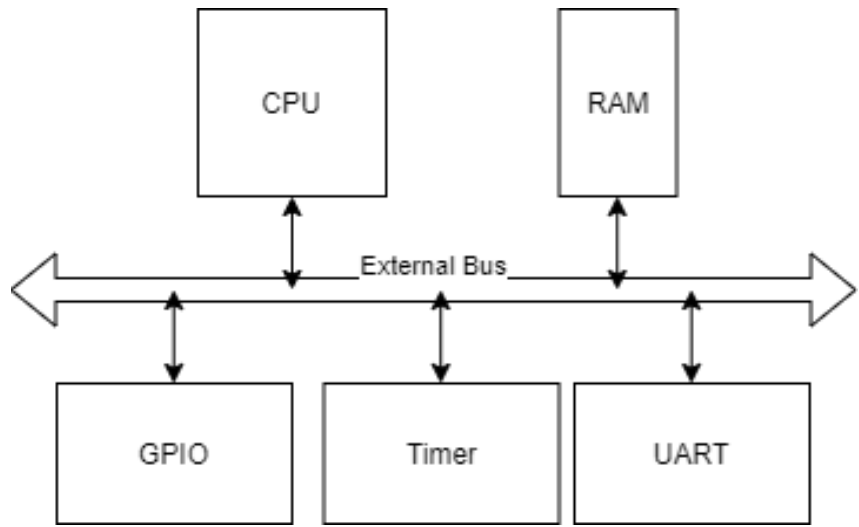

**Figure 2.** The proposed basic SoC architecture.

    A Bus controller IP

With respect to the I/O unit of the SoC, the bus needs to send data from CPU to memory/IO-devices or receive the data from them. The bus controller can be further modified for bus data encoding for special purposes easily. In this paper, the bus controller is simply designed for data transfer without any encoding.

A general-purpose I/O (GPIO) IP

This GPIO Port has 32 pins. The IP can set each pin as input or output individually. If the pin is configured as an output, the status (high/low) can be controlled by CPU read/write machine code. If the pin is configured as an input, the IP can get the specific pin status, and any pin can be set to enable an interrupt function. If more ports are needed, the length of an IP or multiple IPs can be configured in the FPGA.

An Universal Asynchronous Receiver/Transmitter (UART) IP

Both Tx and Rx have 128-byte FIFO buffers, and both Tx and Rx can enable the interrupt function. Currently, only 8 bit can be set as character format (no 7 bit format). For this simple UART IP, the highest baud rate can be set up to any specific value. This UART IP's baud rate can be configured easily (ex. 5M). Compared to the highest baud rate of an off-the-shelf CPU (Max to 115200 or 115K), it can be operated much faster. The number of UART IPs can be also configured freely.

A Timer IP

A simple Timer IP can be set to 16 bit up-counter. An interrupt will be generated when the counter register reaches the set threshold. The timer can be easily extended to be up/down-counter with or without interrupt. Also, the number can be chosen as applications' requirement.

C. The Dynamically Configured ISA Platform

The resources of a FPGA are limited. Since the purpose of the CPU implemented in the FPGA is to control peripheral devices or special functions implemented in the FPGA, the resources occupied by the CPU should be as least as possible. So the remaining resources can be reserved for other functions. For this purpose, we designed a mechanism to dynamically modify the CPU configuration that can further reduce the ISA according to any specific program.

As shown in Figure 3, we design a dynamically configured ISA platform. There are three functions: a reduce ISA(R-ISA) config-generator, an Assembler generator and a CPU generator. The target assembly program and the original ISA are the inputs of the platform.

The R-ISA config-generator generates a R-ISA config-file according to the target program. The target assembly program is optional. This means that if there is no assembly code as input, the R-ISA config-file is the same as the ISA config-file.

The CPU generator generates a R-ISA CPU hardware description according to the R-ISA config-file. Currently, the CPU hardware is described by several Verilog files. The generated Verilog files will be compiled and programmed into FPGA.

The assembler generator generates an assembler according to the R-ISA. When there is an Assembly code input, the R-ISA assembler generates a binary machine code file. And the machine code file will be used by the CPU when power up and is initialized from the on board flash.

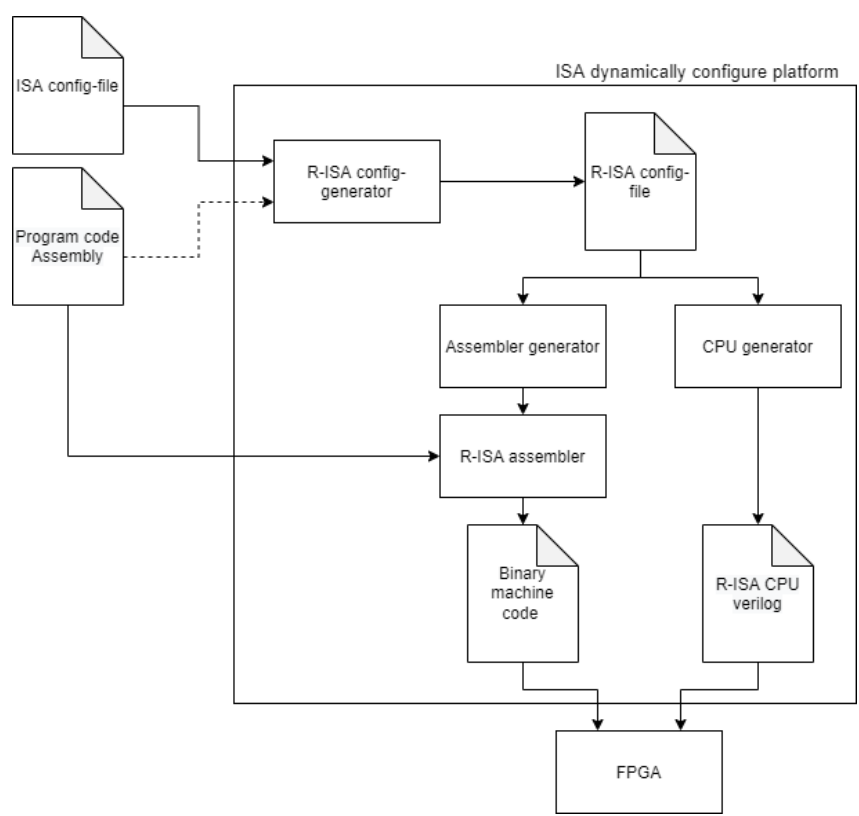

**Figure 3.** Dynamically configured ISA platform flow chat.

**3. Experiment results**

A.  Experiment platform

The FPGA development platform is Intel FPGA Cyclone V GX 5CGXFC9D6F27C7N. It has 301000 Logic Elements(LEs), and 13,917 Kbit/s embedded memory, and 336 I/Os. This paper uses Verilog to design the FPGA function. The development environment is Quartus Prime Editor IDE. The RAM and FIFO used in the experiment are adapted from the On-Chip Memory IP designed by Intel. Currently, the R-ISA generator, the CPU generator, and the assembler generator are designed by C++. The CPU generator generates Verilog descriptions. The Assembler generator generates C++ code which, in turn, is compiled to accept any assembly language based on the R-ISA.

B.  CPU & SoC verification

The following procedures are used for functionality verification of the proposed platform:

　　Instruction cycles simulation

The simulated CPU instruction cycles of the test module is shown in Figure 4. It includes only the proposed CPU and RAM. First, an ISA config-file and a test assembly code are prepared from the platform (Figure 3). The assembly code is assembled into machine code. And the machine code is download into the RAM. The contents of the RAM are shown in Table 1.

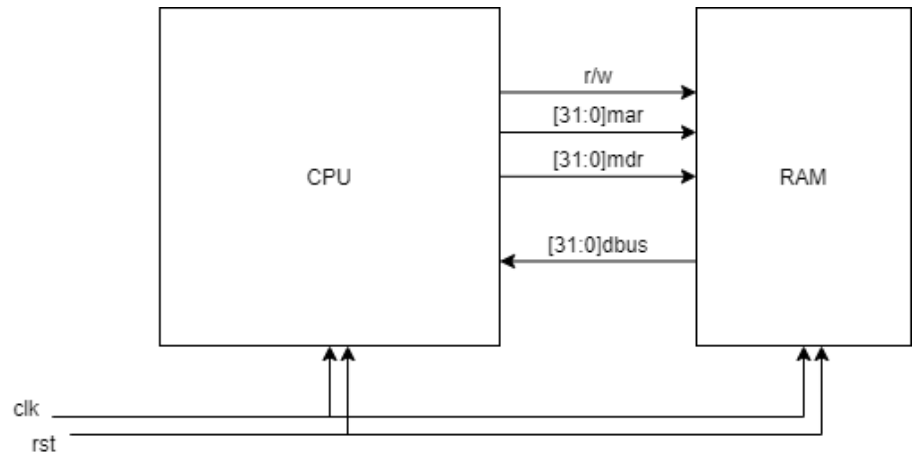

**Figure 4.** Generated CPU instruction cycles test module.

**Table 1.** Generated CPU instruction cycles test assembly code.

| RAM Address | Assembly code | Function | Generated-machine code by platform |
| --- | --- | --- | --- |
| 0x00 | LDI R4, 0x04 | R4 = 0x04 | 0x08400004 |
| 0x04 | LDI R5, 0x08 | R5 = 0x08 | 0x08500008 |
| 0x08 | ADD R8, R5, R4 | R8 = R5+R4 | 0x13854000 |
| 0x0C | ST R8, 0x80 | [0x80] = R8 | 0x01800080 |

These codes (both Verilog code and binary code) are generated by the CPU generator and the R-ISA generator. The simulation (Figure 5) shows the interaction between the CPU and RAM. The simulation shows that the CPU is fetching commands and accessing data. By using the Intel Quartus Prime IDE to create University Program VWF, the built-in waveform simulation tool can perform waveform simulation.

The waveform simulation is shown in Figure 5. At 5ns, the CPU fetches the RAM's address 0x00 (mar) instruction, and the RAM is set to read mode by the CPU (m_rw goes up). After 5ns (at 10ns), the RAM replies (dbus) 0x08400004 (op-code: loading value 0x04 into R4). At 45ns, the CPU finishes the first instruction cycle and starts to fetch the next instruction whose content is at the RAM's address 0x04 which value 0x08500008 (op-code: loading value 0x08 into R5). At 85ns, the CPU fetches next instruction (0x08). This instruction is 0x1385400 (putting the value of add R5 and R4 into R8). And the next instruction is 0x01800080 (store R8 to address 0x80). At 145ns, we can see that the CPU sends value 0x0000000C to the data bus (mdr), and the RAM is set to write mode by the CPU (m_rw goes down). The simulation shows that the CPU has executed 4 instructions (Table 1) correctly. The result shows that the functionality and timing of the CPU when data access to the RAM control are correct.

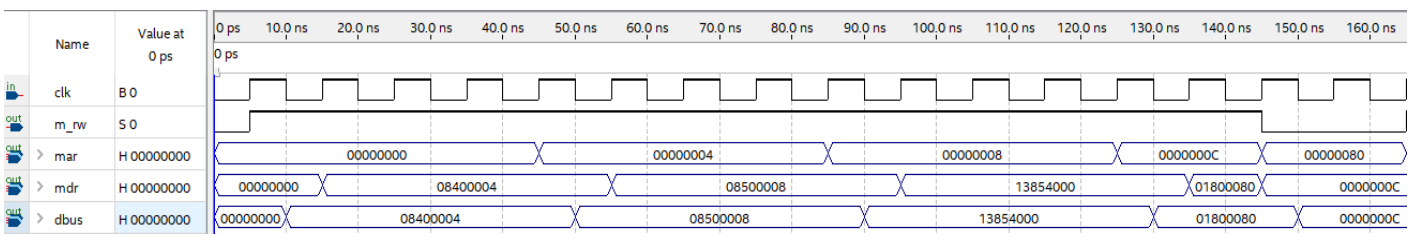

**Figure 5.** Generated CPU instruction cycles verification simulated waveform.

Bus controller & I/O verification

For the function of Bus & IO, same simulation procedure is used for verification. As shown in Figure 6, we write a data to GPIO pin value register at 25ns (mdr) which address is 16388 (mar). The GPIO pin state (gpio_pin) is changed by the data written at 30ns. And we read the GPIO pin value register value at 65ns (m_rw goes up). The GPIO pin value register replies with the correct data (dbus) at 70ns. It shows the Bus can correctly prepare data within 1 clock. For both of the bus controller and I/O devices, the simulation shows that the timing and functionalities are correct.

The correct waveform is preliminarily verified that the CPU, Bus, and IO functions are correct. The next step is the actually verify them on FPGA.

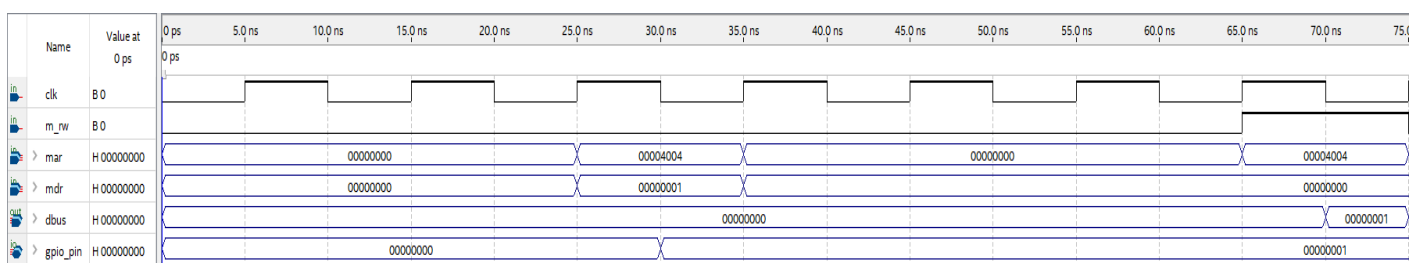

**Figure 6.** Bus controller & I/O verification simulated waveform (GPIO pins value register address is 16388).

The actual FPGA programming and running

This verification is to program a predesigned SoC (Figure 2) function (Figure 7) into the target FPGA platform, and actually test the function. The test result shows the proposed SoC test case can correctly execute all the behaviors of the scenario we designed. As shown in Figure 8, the UART is set to 115200 baud rate. And the UART Tx can correctly send messages every seconds. At the same time, a timer is set to change a GPIO IP. The timer can correctly interrupt every two seconds to toggle the GPIO pin. All the behaviors of CPU, BUS, GPIO, UART, Timer, and RAM are all correctly executed. Therefore, it is confirmed that the CPU & SoC can operate normally. Because of different FPGA quality, the highest clock can be different. This paper uses a 50MHz system clock, and the machine cycle of the CPU is 4 clocks (Table 2).

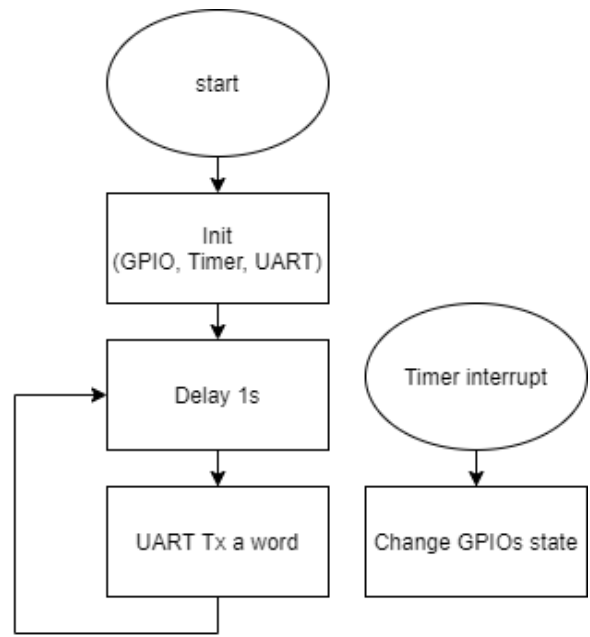

**Figure 7.** SoC actual running test flowchart.

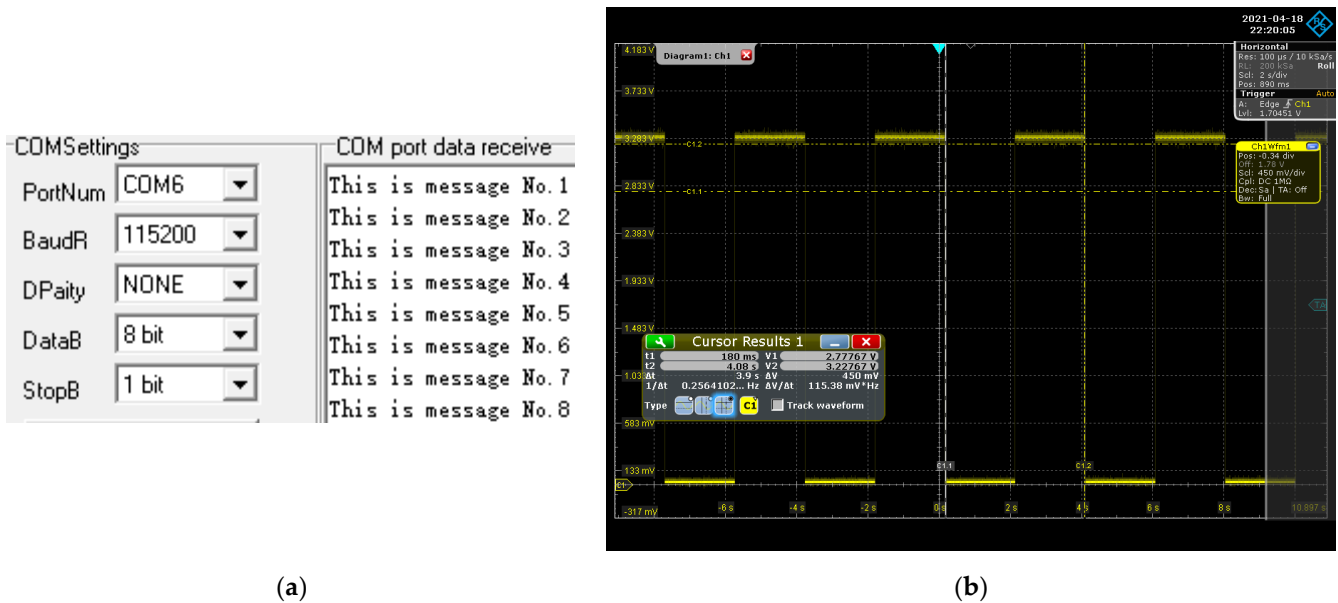

(**a**)                          (**b**)

**Figure 8.** MCU running result. (**a**) UART Tx sends a message to computer every one second; (**b**) Timer interrupts every 2 seconds and toggles the GPIO pin.

**Table 2.** CPU performance.

| Clock frequency | Instruction cycle |
| --- | --- |
| 50 MHz | 4 clocks |

C. Dynamic configuration effect

As the simulation and physical verification are done, the flexibility of the platform is compared with other SoC design platform. The mechanism we proposed can dynamically configure the CPU. And this can automatically remove redundant resource not used by the customized ISA. Here we use MUL (multiplication instructions) and DIV (division instructions) as an experimental example. If the target SoC needs not MUL/DIV of an ISA, the CPU can be configured without multiplier and divider and save the precious FPGA resources. The saved resources can be yielded to other IP functions or reduce FPGA cost. To compare the FPGA resources consumed, the CPU with and without MUL and DIV are shown in Table 3. Although the Register and control unit are the same, the adaptive R-ISA generator scheme can reduce the Look-up Table (LUT) down to 73% of resources.

**Table 3.** FPGA resource comparison on Full-ISA CPU and R-ISA CPU (27% reduction).

| ISA CPU | Used LUTs |
|---|---|
| Full instruction CPU | 4749 (100%) |
| R-ISA CPU (without MUL & DIV) | 3501 (73%) |

D. Demo OS running on MCU

After the reduction scheme is justified, a practical application is used to demonstrate the performance. This demonstration shows a simple operation system (OS) running on the MCU to control an LED matrix (24x32 BS2812) by using the platform we designed (Figure 3). The OS has 2 tasks, one of them is used to control the UART to receive picture pixel RGB values from PC, another task sends the picture data signal to the LED matrix. The PC is running a human face segmentation algorithm [23] given by OpenCV [24]. After the face are cropped, the bit-mapped faces are randomly chosen as the demo target. After receiving the bit-mapped image, the demo target shows the image to the LED matrix. There are 2 tasks and an OS written in the demo. These 2 tasks are co-worked by a circular buffer controlled by the MCU. All the CPU bus controller, memory, Timer, GPIO, UART IPs are auto-generated by the platform. All the OS and tasks are coded in assembly. The binary machine code is generated by the auto-generated R-ISA assembler. Because the hardware and software can be easily customized for highest efficiency, the UART can be configured up to 5M baud rate. As a result, the LED matrix can be successfully controlled at high frame rate. The PC can crop the face (by OpenCV) and transfer the image to the FPGA-MCU to control the LED matrix and display the cropped images at 129 frame per second (Figure 9). The frame rate can be higher if the ISA and assembly program are further improved.

The demonstration shows the power of a customized MCU can be used easily to control hi-speed data transformation and transmission by dedicated and R-ISA with low resource and low MCU clock requirement.

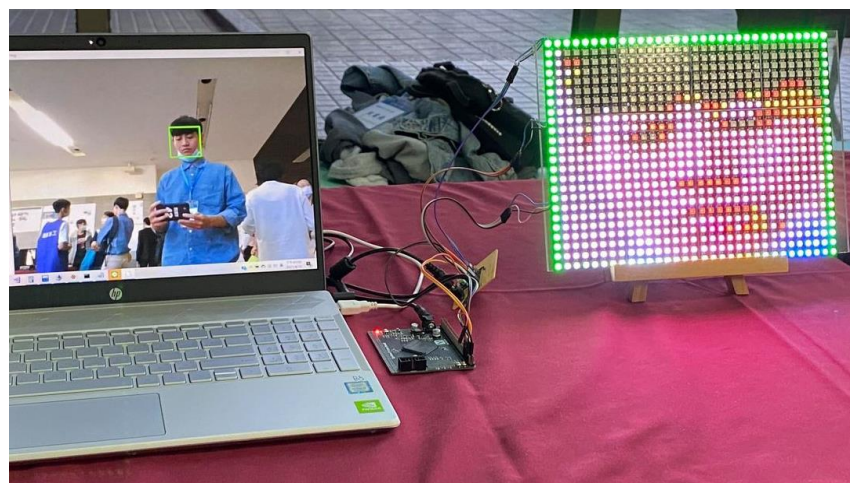

**Figure 9.** FPGA-MCU controlled LED matrixes with OS.

E. Comparison with other CPU

There are many open-source CPU cores available [25,26]. Some of them are very complex and fully supported by software tool-chain. Some of them are still in the very early design stage. But none of them can be so flexible comparing to the proposed architecture. Table 4 lists some of the cores that can be surveyed as a general comparison. From the comparison, ARM cores and RISC-V cores have fully tool-chain support. The flexibility of tool-chain cannot be modified easily. The OP-code modification is not possible for these platform. Contrast to the fixed tool-chain support and OP-code modification inability, the proposed platform has the highest flexibility. The proposed platform can also give enough performance at reduced resource requirement derived from the ISA reduction and the closely related tool-chain and platform auto-generation support.

**Table 4.** Using FPGA resource comparison on this paper and others CPU designed.

| CPU core | ISA Flexibility | Tool chain support | LUT | DSP | Opcode modification |
| --- | --- | --- | --- | --- | --- |
| Arm cores | None | Full | NA | NA | Impossible |
| RISC-V Freedom [27] | Option ISA | Partially or Full | 13062 | 2 | Impossible |
| RISC-V Pulpino RI5CY [28] | Option ISA | Partially or Full | 14616 | 8 | Impossible |
| RISC-V Rocket [29] | None | Partially or Full | 17,144 | 0 | Impossible |
| SuperH J-core [30] | None | Partially or Full | 6589 | NA | Impossible |
| CPU032I [21] | None | Partially | 4498 | 2 | Impossible |
| The proposed full-ISA CPU | Flexibility ISA | Partially | 4749 | 2 | Available |
| The proposed R-ISA CPU | Flexibility ISA | Partially | 3501 | 0 | Available |

## 5. Conclusions

It is quite convenient to use a CPU for product development of various applications. Although its flexibility is very helpful, it is still limited by the CPU instruction set architecture (ISA). The main features of field programmable gate array (FPGA) include flexibility, integration, and cost saving. In this paper we propose a FPGA-based platform that can take advantage of the flexibility to extend the limitation. If an MCU is designed based on FPGA, it would be very flexible on the design issue of CPU ISA, CPU resource requirements, and CPU architecture. Based on the flexible CPU ISA, users can add or reduce new hardware functions and instructions according to their requirements. The dynamic configuration mechanism can remove unused functional hardware during programming to yield precious FPGA resources. This paper also proposes a tool-chain synthesis procedure based on the application. An R-ISA is firstly generated based form the application. And a tool-chain synthesis procedure is applied to generate related hardware/software tools which, in turn, generate the target hardware and software. A case study shows the power of design flexibility: resource requirement is reduced, clock speed remains the same while the image control and transfer performance are faster. The demonstration can be further improved by improving the ISA and application programs. For the application designers, this paper proposes an automatic Verilog & assembler generation mechanism to facilitate the software engineers to use this technology.